\documentclass{ws-ijmpd}

\begin{document}

\markboth{Araudo, Bosch-Ramon \& Romero}
{Jet-Cloud Interactions in AGNs}

%
\catchline{}{}{}{}{}
%

\title{Jet-Cloud Interactions in AGNs}

\author{Anabella T. Araudo}

\address{Instituto Argentino de
Radioastronom\'{\i}a (CCT La Plata, CONICET), C.C.5, (1894)\\ 
Villa Elisa, Buenos Aires, Argentina\\ 
Facultad de Ciencias Astron\'omicas y Geof\'{\i}sicas, UNLP, Argentina\\
aaraudo@fcaglp.unlp.edu.ar}

\author{Valent\'{\i} Bosch-Ramon}

\address{Max Planck Institut f\"ur Kernphysik, Saupfercheckweg 1\\ 
Heidelberg 69117, Germany
vbosch@mpi-hd.mpg.de}

\author{Gustavo E. Romero}

\address{Instituto Argentino de
Radioastronom\'{\i}a (CCT La Plata, CONICET), C.C.5, (1894)\\ 
Villa Elisa, Buenos Aires, UNLP, Argentina\\ 
Facultad de Ciencias Astron\'omicas y Geof\'{\i}sicas, UNLP, Argentina\\
romero@fcaglp.unlp.edu.ar}

\maketitle

\begin{history}
\received{Day Month Year}
\revised{Day Month Year}
\comby{Managing Editor}
\end{history}

\begin{abstract}
Active galactic nuclei present continuum and line emission. The former
is produced by the accretion disk and the jets, whereas the latter is
originated by gas located close to the super-massive black hole. The
small region where the broad lines are emitted is called the
broad-line region. The structure of this region is not well known, 
although it has
been proposed that it may be formed by small and dense ionized clouds
surrounding the supermassive black-hole. In this work, we study the interaction of one
cloud from the broad line region with the jet of the active galactic nuclei. We explore the high-energy
emission produced by this interaction close to the base of the jet.
The resulting radiation may be detectable for nearby non-blazar
sources as well as for powerful quasars, and its detection could give
important information on the broad line region and the jet itself.
\end{abstract}


\section{Introduction}	

Active galactic nucleus (AGNs) are systems mainly composed by a 
supermassive black-hole (SMBH),
an accretion disk, and bipolar relativistic jets. AGNs produce
non-thermal continuum emission along the whole spectrum, from radio
to $\gamma$-rays. The high-energy non-thermal radiation is expected to
come from the jets, which are formed by a magnetized plasma moving
relativistically.  Besides the emission in the continuum, AGNs present
optic and UV lines. Some of these lines are observed with a broad
FWHM, with an associated velocity for the emitting gas of $\sim
10^9$~cm~s$^{-1}$. The region where these lines are formed is called
the broad line region (BLR), and surrounds the SMBH. The size of the
BLR is related to its luminosity, $L_{\rm BLR}$ [\refcite{kaspi}].
For instance, for $L_{\rm BLR}\sim 10^{44}$~erg~s$^{-1}$ (Faranoff-Riley (FR) I
case), $R_{\rm BLR}\sim 6\times 10^{16}$~cm, and for $L_{\rm BLR}\sim
10^{46}$~erg~s$^{-1}$ (FR II case) $R_{\rm BLR}\sim 5\times
10^{17}$~cm. The BLR is thought to be filled with a clumpy medium
composed by cold clouds ($T\sim 10^4-10^5$~K) of radius $R_{\rm c}\sim
10^{13}$~cm [\refcite{rees}]. In this work we study the interaction
between a cloud of the BLR with the relativistic jet close to the SMBH
in an AGN. Assuming standard values of the cloud parameters, and
adopting a hydrodynamical supersonic jet, we estimate the high-energy
emission produced by this interaction for two different kinds of AGN:
a non-blazar source of FR~I type, and a powerful flat-spectrum radio
quasar of FR~II type. 

\section{The physical scenario}

A cloud with a radius $R_{\rm c}$ that moves with a velocity $v_{\rm
  c}=10^9$~cm~s$^{-1}$, completely enters into the jet in  a time
$t_{\rm c}\sim 2\,R_{\rm c}/v_{\rm c}\sim 2\times 10^4$~s. For an
effective penetration of the cloud into the jet, a huge contrast
between the cloud and the jet densities ($\chi=n_{\rm c}/n_{\rm j}\gg
1$) is necessary. The cloud density $n_{\rm c}$ is fixed to $\sim
10^{10}$~cm$^{-3}$ [\refcite{kazanas,tavecchio}] and the density of
the jet, $n_{\rm j}$, is determined through the equation: $L_{\rm
  j}=\sigma_{\rm j}(\Gamma-1)n_{\rm j} m_{\rm p} v_{\rm j} c^2$, where
$v_{\rm j}\sim c$ is the velocity of the jet, $\Gamma\sim 10$ the jet
Lorentz factor, and $\sigma_{\rm j}=\pi R_{\rm j}^2$ is the section of
the jet at the interaction height $z_{\rm int}$, where $R_{\rm j}\sim
0.1 z_{\rm int}$. In order to obtain $n_{\rm j}$ we need to fix  $z_{\rm int}$
at which the cloud penetrates into the jet.

When the cloud penetrates into the jet, a shock is formed and
propagates through the cloud at a velocity  $v_{\rm sh}=v_{\rm
  j}((\Gamma-1)/\chi)^{1/2}$. In a time $t_{\rm cc}\sim 2\,R_{\rm
  c}/v_{\rm sh}$  the shock crosses the whole  cloud. We focus on the
stage when the cloud is inside the jet ($t_{\rm cc}>t_{\rm c}$) at the zero 
order approximation,  implying that the interaction
should take place at least at $z_{\rm int}=2.5\times 10^{15}$ and
$2.5\times 10^{16}$~cm for an FR~I ($L_{\rm j}\sim
10^{44}$~erg~s$^{-1}$) and  an FR~II ($L_{\rm j}\sim
10^{46}$~erg~s$^{-1}$), respectively. At such $z_{\rm int}$, we obtain
for all cases  $n_{\rm j}=1.2\times 10^6$~cm$^{-3}$ and $\chi\sim
10^4$. 

The shock heats the cloud material up to a temperature $T\sim 2\times
10^9$~K. The hot plasma cools via thermal Bremsstrahlung radiation with a
thermal luminosity $\sim 10^{38}$~erg~s$^{-1}$, peaking at soft
$\gamma$-rays.  A bow shock is also formed in the jet, reaching the
steady state at a distance $\sim R_{\rm c}$ from the cloud in a time
$t_{\rm bs}\sim R_{\rm c}/v_{\rm j}\ll t_{\rm cc}$. The cloud could
escape from the jet in a time $t_{\rm j}\sim 2\,R_{\rm j}/v_{\rm
  c}\sim 5\times 10^5$ (FR~I case) and  $5\times 10^6$~s (FR~II case),
being $t_{\rm j}\gg t_{\rm cc}$, although Kelvin-Helmholtz and
Rayleigh-Taylor instabilities can destroy the cloud in a time $t_{\rm
  KH/RT}\sim$ few times $t_{\rm cc}$, shorter than $t_{\rm j}$. We
note that the shocked cloud is also accelerated by the jet and might
reach a velocity $\sim v_{\rm j}$, but it is likely that before this
happens the cloud escapes from the jet.

\section{Non-thermal emission}

In the bow shock, particles can be accelerated via relativistic
Fermi-I mechanism. Given the much lower velocity of the cloud shock,
we will neglect at this stage its role to accelerate particles. The
acceleration rate of particles, for which we adopt here a
phenomenological prescription: $\dot{E}=0.1\,qBc$, depends on the
magnetic field $B$ in the post-shock region of the bow shock 
[\refcite{asterberg}]. We will
consider three different cases, varying the value of $B$ and the
luminosity of the jet: i) case 0, with $L_{\rm j}\sim
10^{44}$~erg~s$^{-1}$ and $B=2.8\times 10^{-2}$~G (FR~I case; dominant
BLR photon energy density); ii) case I, with $L_{\rm j}\sim
10^{46}$~erg~s$^{-1}$ and $B=4.1\times 10^{-3}$~G (FR~II case;
dominant BLR photon energy density); and iii) case II, with $L_{\rm j}\sim
10^{46}$~erg~s$^{-1}$ and $B=1.1\times 10^3$~G (FR~II case; dominant
magnetic energy density, 10\% of equipartition with the post-shock
matter).  The magnetic fields for cases 0 and I are very much below
equipartition. These cases are considered to explore the situation
when external Compton dominates the radiation output, where the
external (BLR) energy density is  $u_{\rm BLR}\sim L_{\rm BLR}/\pi
R_{\rm BLR}^2 c$.  In case II, for $B$-values close to equipartition,
synchrotron emission will dominate. Work on the case when synchrotron
self-Compton is the dominant radiation channel is on-going. 

The maximum energy ($E_{\rm max}$) achieved by electrons is
constrained by the escape of these particles from the bow-shock
region, advected by the shocked material of the jet on a time $t_{\rm
  esc}\sim 3\,R_{\rm c}/c$  (case 0: $2\times 10^{13}$~eV and case I:
$4\times 10^{12}$~eV), and by synchrotron losses (case II: $5\times
10^{11}$~eV). 
For protons the maximum energy is determined by the size of the
acceleration region, $\sim R_{\rm c}$ (case 0: $8\times 10^{13}$~eV,
case I: $10^{13}$~eV, and case II: $3\times 10^{18}$~eV). 

We assume here that the 20\% of the $L_{\rm j}$ fraction transferred
to the cloud is converted into (non-thermal) luminosity of the
accelerated particles, i.e. $L_{\rm NT}=0.2\,(\sigma_{\rm
  c}/\sigma_{\rm j})\,L_{\rm j}$.  We determine the constant $K_{e,p}$ of
the injection function $Q$, assuming a power-law with an index
$p=-2.2$ and a cut-off at higher energies:
$Q_{e,p}=K_{e,p}\,E^{-p}\,e^{-E/E_{\rm max}}$.  The electron energy
distribution, $N_e$, is determined by the escape of particles  and
radiation losses (synchrotron and inverse Compton (IC)), reaching the steady state on a
time $\ll t_{\rm cc}$. The steady distribution of these relativistic
leptons is $N_e=Q_e\,t_{\rm esc}\propto E^{-p}$ (in cases 0 and I;
escape dominance) and $N_e=Q_e\,t_{\rm synch}\propto E^{-p-1}$ (in case II;
synchrotron cooling).  On the other hand, protons escape from the
bow-shock region after losing a negligible fraction of their energy by
$pp$ interactions.  


As noted above, the most important emission process is IC scattering
in cases 0 and I.  In the former, the achieved luminosities are
$L_{\rm synch}\sim 10^{36}$ and $L_{\rm IC}\sim 10^{38}$~erg~s$^{-1}$ (peaking 
in hard X-rays and in $\gamma$-rays, respectively), and in the latter, $L_{\rm synch}\sim
10^{34}$ and $L_{\rm IC}\sim 10^{37}$~erg~s$^{-1}$ (peaking in soft 
X-rays and in $\gamma$-rays, respectively). In case II, corresponding 
to the larger value
of $B$, the synchrotron emission is the most important channel of
energy loss,  peaking in the the infrared and decreasing smoothly
afterwards up to soft $\gamma$-rays.  In this case, synchrotron and IC
luminosities reach values $\sim 10^{39}$ and $10^{32}$~erg~s$^{-1}$,
respectively. The IC maximum is always in the sub-TeV range, given the
strong photon-photon absorption in the BLR photon field, producing
pairs that will also generate synchrotron and IC radiation
(e.g. [\refcite{aharonian}]). Regarding relativistic protons, since
the density of particles in the bow-shock region is much lower than that of 
the cloud density, $pp$ interactions are not an effective radiative
process. For this reason, the most energetic protons ($E_p\sim E_{\rm
  max}$) can reach the cloud via diffusion before being advected away
from the bow-shock region. Inside the cloud, the relativistic protons
lose energy via $pp$ collisions, yielding (absorbed) $\gamma$-ray luminosities
$L_{pp}\sim 10^{36}$, $10^{38}$ and $10^{39}$~erg~s$^{-1}$ for the
cases 0, I and II, respectively. Case II is an interesting one since
radiation above 100~TeV may be detectable. We note that secondary
leptons and neutrinos with luminosities similar to those of $\gamma$-rays,
would be also produced inside the cloud due to $pp$
interactions, and their study will be carried out in future work.

\section{Final remarks}

The relation flux/luminosity/distance:  $F\sim 10^{-12} (L_{38}/d_{\rm
  Mpc}^2)$~erg~s$^{-1}$~cm$^2$,
and the luminosity values given above, show that one-cloud/jet
interaction fluxes predicted in the present contribution could be 
detectable at X- or
$\gamma$-rays up to distances of $\sim 10$~Mpc. The interaction of a
cloud at the adopted $z_{\rm int}$ could lead either to persistent but variable
emission if there were many clouds interacting with the jet [\refcite{owocki}],
or to sporadic emission with a certain duty cycle if clouds interact with
the jet  from time to time. For the parameters adopted here,
the jet-cloud interaction would be a persistent activity if the number
of clouds is $N_{\rm c}>10^8$ (for an FR~II and assuming a life time of a
cloud in the jet of $\sim t_{\rm j}$), and sporadic if $N_{\rm c}$
is smaller. In the former case, the actual luminosities should be
obtained here multiplying by $(N_{\rm c}/10^8)$, and in the
latter, the interaction duty cycle would be proportional to $100\times
(N_{\rm c}/10^8)$\%. 
Notice that at larger
$z_{\rm int}$, the emission will be reduced even if a larger number of
clouds could penetrate the jet, given the smaller cloud to
jet section ratio. Finally, we remain that given the properties of the 
emitter, the jet-cloud interaction radiation is not beamed and 
almost isotropic.

\section*{Acknowledgments}

A.T.A. thanks the Max Planck Institut f\"ur Kernphysik for its kind
support and hospitality.  G.E.R. and V.B-R.  acknowledge support by
DGI of MEC under grant AYA2007-6803407171-C03-01, as well as partial
support by the European Regional Development Fund
(ERDF/FEDER). V.B-R. gratefully acknowledges support from the
Alexander von Humboldt Foundation.

\end{document}